\documentclass[11pt,a4paper]{article}
\usepackage{epsfig,amsmath,amssymb,array,dcolumn,subfigure,rotating}
\textwidth 
15 true cm

\begin{document}

\newpage

{\large \bf \center
    {Osmotic properties of polyethyleneglycols:} \\
    {quantitative features of brush and bulk scaling laws}   \\[3mm]
}
{\center Per Lyngs Hansen,$^{\dagger \star}$ Joel A. Cohen,$^{\dagger 
\ddagger}$
Rudi Podgornik,$^{\dagger \S}$\\
and V. Adrian Parsegian$^{\dagger}$ \\[3mm]
{$^{\dagger}$Laboratory of Physical and Structural Biology} \\
{National Institute of Child Health and Human Development} \\
{National Institutes of Health, Bethesda, MD 20892-5626}
\\[3mm]
{$^{\star}$MEMPHYS - Center for Biomembrane Physics} \\
{University of Southern Denmark} \\
{Campusvej 55, DK-5230 Odense M} \\
{Denmark} \\[3mm]
{$^{\ddagger}$Department of Physiology, University of the Pacific} \\
{San Francisco, CA 94115} \\[3mm]
{$^{\S}$Department of Physics} \\
{Faculty of Mathematics and Physics} \\
{University of Ljubljana, SI-1000 Ljubljana, Slovenia} \\[5mm]
}
\vspace{1cm}

\vspace{1cm}
\noindent{Running title:} PEG brush and bulk
scaling laws

\date{\today}
\newpage
\begin{center} {\large \bf Abstract} \end{center}

\vspace{3mm}
\noindent
From glycosylated cell surfaces to sterically stabilized 
liposomes, polymers attached to membranes attract biological and 
therapeutic interest.  Can the scaling laws of polymer "brushes" 
describe the physical properties of these coats?  We delineate conditions 
where the Alexander - de Gennes theory of polymer brushes successfully 
describes the intermembrane distance vs. applied osmotic stress data 
of Kenworthy et al. for PEG-grafted multilamellar liposomes [Biophys. J. 
(1995) 68:1921].  We establish that the polymer density and size in the brush must be high enough that, in a bulk solution of equivalent density, the polymer osmotic pressure is independent of polymer molecular weight (the des Cloizeaux 
semi-dilute regime of bulk polymer solutions).  The condition that attached polymers behave as semi-dilute bulk solutions offers a rigorous criterion for brush scaling-law behavior.  There is a deep connection between the behaviors of polymer solutions in bulk and polymers grafted to a surface at a density such that neighbors pack to form a uniform brush.  In this 
regime, two-parameter unconstrained fits of the Alexander - de Gennes brush 
scaling laws yield effective monomer lengths of 3.3 to 3.5 \AA, which agree with structural predictions.  The fitted distances between grafting sites are larger than expected from the nominal content of PEG-lipids; the chains apparently saturate the surface.  Osmotic stress measurements can be used to estimate the 
actual densities of membrane-grafted polymers. 

\newpage
\noindent
{\bf INTRODUCTION}

\vspace{5mm}
\noindent
Membrane surfaces decorated with end-grafted polymers are
ubiquitous in biology (for example, glycosylated cell surfaces such
as the red blood cell glycocalyx).  Organelles (e.g., microtubules)
also posses polymeric "hairs" that are integral to their function
(Sackett, 1995).  A relatively well-defined example of a
surface-attached polymeric coat is that of poly(ethylene glycol)
(PEG), also known as poly(ethylene oxide) (PEO).  These polymer 
coatings attract strong interest because of their ability to shield
surfaces from close interactions with other surfaces including
macromolecules.  They thus provide steric stabilization to colloidal
suspensions (Lasic and Martin, 1995), biocompatibility to
medically-implanted materials (Dumitriu, 2002), and
"stealth" properties to intravenously-injected liposomes.  Such
liposomes are in current therapeutic use as vehicles for
{\sl in-vivo} drug delivery (Lasic and Martin, 1995).

It is commonly believed that in order to provide effective shielding 
of a surface
from interactions with aqueous substances, the attached polymers must
provide a surface layer of adequate coverage and thickness, i.e., they
must approximate a surface "brush" (Szleifer, 1997).  Polymers dilutely
grafted to a surface are said to form "mushrooms" when the mean
distance between grafting sites $D$ is larger than the polymer size $R_F$
(Flory radius), so that individual polymer chains remain widely
separated and do not interact with each other.  If $D$ is decreased
(higher grafting density) and/or $R_F$ is increased (longer polymers)
such that $D \sim R_F$, then individual chains start to overlap and
the polymers begin to interact.  This "overlap criterion" defines the
so called "mushroom-to-brush transition", and is commonly used to
signify the beginning of brush-like behavior of the grafted polymer
layer.  A major point of this paper is that brush scaling laws are not
applicable in this so-called "weak overlap regime".  Validity of brush
scaling laws can be expected only for {\it strong} overlap, which is
defined by a criterion different than the above.

The physicochemical characterization of a system of PEG chains 
end-grafted to the surface
of a supporting lipid bilayer has biological and practical 
significance, but is also relevant to polymer physics (de Gennes, 
1987).  A well-tested scaling theory of the polymer brush, due to
Alexander and de Gennes (AdG) exists and is currently well accepted 
(Alexander, 1977; de Gennes, 1987).  It is thus important to 
ascertain whether, or under what conditions, the grafted PEG system 
adheres to established brush scaling laws.

To date, several investigators have addressed this issue by 
performing compression experiments in which two or more PEG-grafted 
bilayers are forced together, and force vs. intermembrane distance is 
measured.  The geometry of such an arrangement is illustrated in Fig. 1,
where parameters relevant to the theoretical analysis (see below) are also 
shown.  For such data to be relevant to brush scaling
laws, two criteria must be met: (1) The intermembrane forces must be 
dominated by interactions
between the polymer chains, as opposed to electrostatic, van der 
Waals, hydration, or undulation  forces.  (2) The grafted polymer 
chains must exist in a strong brush conformation, as
opposed to mushrooms or a weak brush-like state not far from the 
mushroom-to-brush transition.

Kuhl et al. (1994), using a surface force apparatus (SFA), measured 
force vs. intermembrane distance for mica-supported bilayers whose 
apposed surfaces were composed of mixtures of 
distearoylphosphatidylethanolamine (DSPE) and PEG-grafted DSPE.  
Long-range repulsive forces were observed, and subtraction of the 
electrostatic component yielded repulsions attributable to 
interactions between the PEG surface layers.  The densest conditions 
studied were PEG-2000 (45 monomers) at a grafting density 
corresponding to 9 mol \% PEG-PE.

Kenworthy et al. (1995a), using osmotic stress (OS), measured osmotic 
pressure vs. intermembrane distance for multilamellar liposomes 
composed of mixtures of
distearoylphosphatidylcholine (DSPC) and PEG-grafted DSPE.  Clear
evidence was given for interactions mediated by the PEG chains.
Nominal densities of the grafted PEGs (for Avanti lipids) ranged up
to 30 mol \% PEG-2000 and 20 mol \% PEG-5000 (113 monomers).

Another technique for measuring intermembrane force vs. distance with 
PEG-grafted liposomes is that of micropipet manipulations (Needham et 
al., 1992; Evans et al., 1996).  However, the conditions for these 
experiments have thus far not been amenable to brush scaling analysis.

In cases where results were analyzed in terms of the AdG brush 
scaling theory (Kuhl et al., 1994; Kenworthy et al., 1995a; Lasic, 
1997), agreement between theory and experiment was not satisfactory. 
However, it has been noted (Szleifer, 1996) that most of the above 
data lies in a broad mushroom-to-brush transition region, and since scaling 
laws do not apply to this intermediate regime, quantitative fits are 
not to be expected.  Szleifer (1996) has discussed the successful 
application of computer simulations and molecular theories to experimental 
data in the mushroom-to-brush transition regime.  
While such approaches apply also in the strong-brush limit, they do not 
yield scaling laws and lack the robust predictions of scaling theory for 
semidilute solutions.  Our interest here is to ascertain whether 
bulk and surface-grafted PEG systems can fulfill the criteria for 
analytical scaling theories, and if so, whether the scaling laws can be 
successfully applied.

The unanswered questions remain:  can the grafted PEG layers on lipid 
membrane surfaces exist in a brush regime, and if so, do they obey brush 
scaling laws?

In this paper we first examine the relation between brush scaling laws 
and the behavior of polymer chains in bulk solution.  Bulk PEG solutions 
are shown to exhibit scaling behavior under experimentally-realizable conditions.  We then examine the validity of AdG scaling laws as applied to 
PEG-grafted lipid bilayers.  An operational criterion is presented for 
determining the PEG brush scaling regime, and the range of 
validity of the scaling laws is shown to be more restrictive than often 
supposed.  When 
applied to those data of Kenworthy et al. that satisfy the brush criterion, 
brush scaling laws are found to be valid.  Further, fits to the data 
yield hitherto difficult-to-obtain information on the density of PEG 
grafts present on the bilayer surface.  As a function of mol fraction 
of added PEG-lipid, these densities plateau at values smaller than the 
nominal densities, indicative of surface saturation 
effects.  The saturation mol fractions of PEG lipids in the bilayer 
are consistent with earlier estimates of this phenomenon.

%---------------------------------------------------------------------------------
\vspace{5mm}
\noindent
{\bf RESULTS}

\vspace{5mm}
\noindent
When flexible polymers such as PEG are end-grafted to a surface,
brush formation begins when $D\simeq R_{F}$, i.e., when the
average distance between grafting sites $D$ is comparable to the Flory
radius $R_{F}=aN^{3/5}$, where $a$ is the effective monomer length,
and  $N$ is the number of monomers per polymer chain.  However, it is
important to realize that a brush is not fully developed unless the monomer
density $\bar{\phi}$ is large enough (and chain overlap is strong
enough) that a semi-dilute solution is formed.  For a description of
semi-diluteness in a brush, see the Appendix and de Gennes (1979).
It is important to emphasize that in a non-compressed brush, the
celebrated linear relation between the brush thickness and number of
monomers $L_{0}\simeq a N (a/D)^{2/3} $, as well as the
molecular-weight-independent relation $\bar{\phi_{0}}\simeq
{(a/D)}^{4/3}$ between the spatially-constant volume fraction of
monomers{\footnote{The Alexander - de Gennes model may be improved in
several respects.  For instance, one may self consistently calculate
an improved monomer density profile, as in Milner-Witten-Cates theory
(Milner et al., 1988).  Such improvements affect our force and osmotic
pressure predictions to a very limited extent (Kenworthy et al.,
1995a) and we shall ignore them here.}} and $D$, are both
consequences of semi-dilute solution behavior in brushes (Alexander,
1977; de Gennes, 1987).

The natural way to test for possible semi-dilute solution behavior in
PEG-lipid structures is to consider PEG under bulk conditions.  We
argue that flexible {\it end-grafted} PEG chains will form
semi-dilute solutions only if PEGs of comparable density form
semi-dilute {\it bulk} solutions.  In bulk solutions one may check
for the desired property by determining
whether the osmotic pressure $\Pi$ is related  to the bulk monomer
volume fraction $\phi$ in the manner predicted by des Cloizeaux (de
Gennes, 1979):
\begin{equation}
\Pi = \alpha (k_{B}T/a^{3}){\phi}^{9/4} ,
\end{equation}
where $\alpha$ is a constant ${\cal{O}}(1)$.  In Fig.~2 we have 
plotted the room-temperature bulk osmotic pressure as a function of 
bulk monomer volume fraction for PEG polymers of various molecular 
weights between 1000 and 20000.  The data were obtained from Rand 
(2002).  Additional data consistent with those shown here are also
available (Reid and Rand, 1997).  

Fig.~2 illustrates three points not previously addressed in studies of bulk PEG osmotic
properties, including virial-expansion (Cohen and Highsmith, 1997) 
and excluded-volume (Reid and Rand, 1997) analyses of PEG osmotic pressures: {\it (i)} At high densities, the osmotic pressures of PEGs with molecular weight exceeding $\sim$1500 
Da indeed approach the des Cloizeaux result Eq.~1, if we take 
$a=3.5$ {\AA} (Kenworthy et al., 1995a) and fit $\alpha = 0.8$. 
Thus, sufficiently long PEG chains under bulk conditions indeed form 
semi-dilute solutions.  {\it (ii)} A crossover from ideal-gas 
behavior $\Pi=(k_{B}T/a^{3})\phi/N$ at low volume fractions to des 
Cloizeaux behavior at high volume fractions takes place at higher 
concentrations the lower the molecular weight.  Moreover, the chain 
overlap condition  $\phi\simeq  \phi^{*}=N^{-4/5}$ (de Gennes, 1979) 
does {\it not} provide a sufficient criterion for the attainment of
des Cloizeaux behavior.  {\it (iii)} PEG-2000 solutions are in the 
scaling regime when the volume fraction is larger than 
$\phi^{\#}_{2000}\simeq 0.15$, rather than the overlap concentration
$\phi^{*}_{2000}=0.05$.  Similarly, in view of the experimental 
uncertainty, PEG-5000 solutions are not in the scaling regime until 
the volume fraction is larger than $\phi^{\#}_{5000}\simeq
0.07-0.09$, rather than $\phi^{*}_{5000}=0.02$.

Bulk-solution analysis of PEG data helps us establish whether AdG theory
applies to stressed PEG-liposomes (Kuhl et al., 1994; Kenworthy et
al., 1995a).  Specifically, just before two brush-covered surfaces are
brought into first contact, $\Pi\simeq 0$ and the thickness of the
layers is $L\simeq L_{0}$.  The volume fraction of monomers is
therefore ${\bar{\phi}}_{0}\simeq (a/D)^{4/3}$.  If we assume $a=3.5$
{\AA} and use the relation $D\simeq {(A/f)}^{1/2}$, where $f$ is the
mol fraction of PEG-lipids and $A$ is the area per DSPC lipid $=48$
\AA ${^{2}}$ (Kenworthy et al., 1995a), we find that solutions of
DSPC:PEG-2000 reach the critical value $\phi^{\#}_{2000}=0.15$ at mol
fraction $f^{\#}_{2000}=0.23$.  Similarly, solutions of DSPC:PEG-5000 reach
the critical value $\phi^{\#}_{5000}=0.07-0.09$ mol fraction at
$f^{\#}_{5000}=0.07-0.10$.

In Fig.~3 we display the Avanti DSPC:PEG-5000 OS data reported by
Kenworthy et al. (1995a).  We focus on the behavior at the highest
coverages corresponding nominally to $f=0.1$ and $0.2$, where bulk
analysis leads us to believe that AdG theory may well apply.

According to AdG theory (Alexander, 1977; de Gennes, 1987) (see
Appendix) the formation of a brush is a compromise between
excluded-volume repulsions, which lead to an
osmotic contribution similar to the des Cloizeaux result, and
grafting constraints that are responsible for entropic elastic
tensions.  The expression for the osmotic pressure reflects the
difference between these effects:
\begin{equation}
      \Pi(L) = \alpha \frac{k_{B}T}{D^{3}}
        \left[ \left({\frac{L_{0}}{L}}\right)^{9/4} -
\left({\frac{L}{L_{0}}}\right)^{3/4} \right]
\end{equation}
where
\begin{equation}
      L_{0} = a N \left( \frac{a}{D} \right)^{2/3}~.
\end{equation}
We conjecture that $\alpha =0.8$ as for bulk PEG solutions (see Eq.~1).

One may test AdG theory by substituting  $ D(L_{0},a)$ from Eq.~3
into the expression $\Pi(L)$ in Eq.~2, and performing
unconstrained fits with the two free parameters $a$ and $L_{0}$.  The
solid lines in Fig.~3 are the results of such fits.  The quality of
the procedure is gauged by the degree to which roughly constant and
"reasonable" values of $a$ are obtained.  We indeed get nearly
constant $a$'s close to the values cited in the literature (Kenworthy
et al., 1995a): $a_{0.1}=3.56$ {\AA} and $a_{0.2}= 3.30$ {\AA}.
Taking $a=3.5$ {\AA}, the fitted values for the brush thicknesses are
$L_{0}=105$ {\AA} for $f=0.1$, and $L_{0} =109$ {\AA} for $f=0.2$, 
respectively.
The uncompressed brush thickness $L_{0}$ varies as expected: the higher
the coverage, the thicker the brush.  However, the grafting densities
implied by these fits are systematically lower than the nominal
values.  Assuming, as before, $f\simeq A/D^{2}$ with $A= 48$
\AA$^{2}$, we find $f_{0.1}=0.07$ and $f_{0.2}=0.08$.

In summary, good fits producing nearly constant (unconstrained)
values of $a$, with fitted $f$'s lower than nominal values but still
within the semi-dilute regime ($f$'s $>f^{\#}_{5000}$), indicate
that scaling analysis of the PEG-5000 data of Kenworthy et al. is
valid and self-consistent.  Saturation effects in these dense brushes
are indicated by lower than nominal values of $f$.

For DSPC:PEG-2000, AdG analysis shows similar saturation effects.  In
the case of 30 mol \% PEG-lipid, fits yield $f_{0.3}<f^{\#}_{2000}$,
thus the monomer density is below the onset of semi-dilute behavior.
Therefore in this case the fitting procedure is not self-consistent.
AdG theory is even less applicable for smaller mol fractions of
PEG-2000 lipids.

\vspace{5mm}
\noindent
{\bf DISCUSSION}

\vspace{5mm}
\noindent
This paper emphasizes that brush and bulk scaling laws are related
and illustrates how this relation yields information about brush
formation in PEG-grafted liposomes.  On one hand, a scaling analysis of
bulk osmotic pressure data for PEGs of various molecular weights
establishes criteria stronger than the popular overlap criterion
$D=R_{F}$ (Kuhl et al., 1994; Kenworthy et al., 1995a; Belsito et
al., 2000) for invoking AdG theory of brush behavior:  {\it In
order to invoke Eq.~2, it must be established that the polymer size
and density in the compressed brush are in a regime where bulk des
Cloizeaux scaling applies.}
Consequently, we are forced to forego AdG analysis for PEG-grafted
liposomes containing PEGs of molecular weight 2000 or less.  On the
other hand, having established that semi-dilute scaling regimes are
within reach for long bulk PEG polymers, we are increasingly
confident that AdG theory (or improvements thereof - see Footnote 1)
is the correct framework for analyzing densely-grafted liposomes
containing PEGs of molecular weight greater than $\sim$2000.

Other experimental studies lend support to the view that long
polymers end-grafted to non-adsorbing substrates under good solvent
conditions have bulk scaling behavior as predicted by des Cloizeaux,
and brush scaling behavior as described by Alexander and de Gennes
(Auroy et al., 1991; Taunton et al., 1990).  Our analysis
illustrates that brush formation with long water-soluble polymers is
fundamentally no different than brush formation with synthetic
polymers in organic solvent.

The fact that our two-parameter fits to the AdG expression (Eqs.~2
and 3) are of such  quality that they lead to unbiased prediction of
fairly constant effective monomer lengths $a\simeq 3.5$ \AA, in good
agreement with structural predictions (Kenworthy et al., 1995a),
leads us to conjecture that osmotic-stress (OS) measurements
constitute a method
for non-trivial, semi-quantitative structure determination of
PEG-grafted liposomes.  Hitherto few physicochemical methods (Belsito
et al., 2000; Montesano et al., 2001) have been invoked to refine
estimates of grafting densities beyond nominal values.  Additional
methods are desireable, not the least because of the technological
importance of PEG-grafted liposomes.  With OS analysis, seemingly
reliable determination of the brush monomer density $\bar{\phi}_{0}$
is possible, assuming of course that reliable OS data can be
obtained in the semi-dilute regime.

Assuming the area per lipid, $A$, to be well-determined, the mol
fraction of PEG-lipids
$f=f(\bar{\phi}_0)$ can be determined too.  Several values have been
reported for the area per lipid of DSPC  (Kenworthy et al., 1995a;
Rand and Parsegian, 1989; Lis et al., 1982). Irrespective of which 
value is used in the calculations, we predict that densely-grafted 
DPSC:PEG-lipid liposomes are subject to surface
saturation effects.  For example, for DSPC:PEG-5000 with nominal mol
fractions $0.1$ and $0.2$, the difference between the fitted
grafting densities is small.  Assuming $A=48$ {\AA}$^{2}$ (Kenworthy
et al., 1995a), $f_{0.1}=0.07 $ and $f_{0.2}=0.08 $.  The origin of
such saturation effects is not yet well understood.  On theoretical
grounds (Hristova and Needham, 1995), one expects that addition of
extra PEG-lipids will cause an increase in lateral pressure and a
decrease in lipid packing area until formation of non-bilayer
structures such as micelles is favored.  In PEG-grafted liposomes with
dipalmitoylphosphatidylcholine (DPPC) as the host lipid, this
mechanism seems to be active (Belsito et al., 2000; Montesano et al.,
2001).  In DSPC:PEG liposomes below the chain-melting temperature the
situation is more complicated.  For DSPC:PEG-5000, Kenworthy et al. 
(1995b) suggest that before micelle formation sets in, a moderate 
increase in PEG-lipid content is accompanied by a transformation from 
an $L_{\beta'}$-like phase to an $L_{\beta}$-like phase.
Interestingly, the transformation between these phases is argued to
set in at mol fractions of PEG-lipid corresponding approximately to
the saturation limit calculated here, i.e., $f_{5000}\simeq 0.1$.

\vspace{5mm}
\noindent
{\bf CONCLUSIONS}

\vspace{5mm}
\noindent
Sufficiently long PEG polymers in bulk solution may form semi-dilute
solutions with
osmotic properties as foreseen by des Cloizeaux (de Gennes, 1979).
Sufficiently dense and thick "brushes" of PEG-lipids end-grafted to
lipid bilayers are therefore expected to behave as confined
semi-dilute solutions with a scaling structure as predicted by
Alexander and de Gennes (Alexander, 1977; de Gennes, 1987).  The
approach to bulk semi-dilute behavior is slow for PEGs of molecular
weight less than several $1000$s.  When attached to lipids, those
PEGs will form brushes satisfying AdG theory only if they are
rather dense with grafting densities exceeding $f^{\#}\simeq 0.2$.  In
practice this condition may be difficult to realize.  Of all the data
reported in the literature (Kuhl et al., 1994; Kenworthy et al.,
1995a) only the osmotic stress data for DSPC:PEG-5000 with nominal
grafting densities in excess of $f^{\#}\simeq 0.10$ approximately
satisfy the AdG theory.  Two-parameter unconstrained fits using the AdG
prediction for osmotic pressure vs. bilayer separation yield good fits
with an effective monomer length $a \simeq 3.5$ {\AA} in agreement with
structural predictions.  The coverages inferred from the fits
are lower than the nominal coverages, an indication of surface
saturation effects in PEG-liposomes which are now beginning to be
understood.  We conjecture that osmotic-stress measurements form
the basis for semi-quantitative structure determinations of
PEG-grafted-liposome surfaces.

A quantitative characterization of brush scaling behavior and
structure relies on a precise identification of semi-diluteness that
is more rigorous than the simple, and often-used,
chain-overlap criterion, $D\simeq  R_{F}$.

\vspace{1cm}
\noindent
{\bf Acknowledgments}

\vspace{0.5cm}
\noindent
We thank Dr. A. Kenworthy for valuable discussions.  One of us (PLH)
would like to thank Prof. F. Pincus for valuable comments.

\newpage
\noindent
{\bf APPENDIX}

\vspace{5mm}
\noindent
{\bf Isolated, non-compressed brush:} (de Gennes, 1987; de Gennes,
1979)
\noindent
In a fully-developed non-compressed brush, chain overlap is
so strong that a semi-dilute
solution of spatially-constant monomer volume fraction
$\bar{\phi}_{0}$ is formed (however, see Footnote 1).  We recall that
under semi-dilute
solution conditions  {\it (i)} the relevant degree of freedom is a
"blob" characterized by its size $\xi(\bar{\phi}_{0})$ and its free
energy $k_{B}T$;  {\it (ii)} the gas of blobs is non-interacting, but
the chains inside a blob interact solely via excluded-volume
repulsions so the number of
monomers $g_{\xi}$ inside a blob is related to blob size by the
Flory relation $\xi(\bar{\phi}_{0})= a g_{\xi}^{3/5}$, where $a$ is
the effective monomer length; and {\it (iii)} most physical
properties become molecular-weight independent.

In non-compressed brushes the blob size is determined by the distance
between grafting sites, $\xi(\bar{\phi}_{0})\simeq D$.  At the same
time, molecular-weight independence of physical properties in
semi-dilute solutions implies that $g_{\xi}= g_{D} \simeq
\bar{\phi}_{0}^{-5/4}$, and $\xi({\bar{\phi}}_{0})\simeq a
\bar{\phi}_{0}^{-3/4}$.  It follows that $\bar{\phi}_{0}\simeq
(a/D)^{4/3}$.  Polymer chains in the brush form strings of blobs.
The length of a non-compressed string is the blob size times the
number of blobs, i.e.,  $L_{0}\simeq D\times (N/g_{D}) \simeq
aN(a/D)^{2/3}$.  This well-known linear relation between $L_{0}$ and
$N$ reflects the strong stretching of chains in a brush and is
largely a consequence of semi-dilute solution behavior in the brush.

\vspace{5mm}
\noindent
{\bf Compressed brush:}
\noindent
It is instructive to view the formation of a brush as a
compromise between excluded-volume monomer repulsions, which lead to
an osmotic contribution, and confinement effects (due to the grafts)
that are responsible for entropic
elastic tensions.  In the absence of osmotic stress these effects
balance each other.
When subjected to compression, apposing brushes begin to overlap,
the brush thickness $L$ decreases, and the monomer density in the
brush increases: $\bar{\phi}_{L} =\bar{\phi}_{0}\times (L_{0}/L)$.
The osmotic stress generates an imbalance between osmotic and elastic
terms which may be described as compression of a string of blobs.  In
fact, the resulting osmotic pressure can be derived from the free
energy per chain $F_{c}= F_{os} + F_{e}$, where the osmotic term is
$F_{os}\simeq k_{B}TN{{\bar{\phi}}_{L}}^{5/4}$, namely $k_{B}T$ per
blob times the number of blobs, as in bulk semi-dilute solutions.  The
elastic term is a Flory-type entropic elastic free energy for an ideal
random walk of blobs, $F_{e}\simeq
k_{B}TL^{2}/R^{2}(\bar{\phi}_{L})$, with ideal radius squared given by
the blob size squared times the number of blobs,
$R^{2}(\bar{\phi}_{L})=\xi^{2}(\bar{\phi}_{L})\times (N/g_{L})$.  If
we invoke the Alexander conditions $g_{L}\simeq
\bar{\phi}_{L}^{-5/4}$ and $\xi({\bar{\phi}}_{L})\simeq
a\bar{\phi}_{L}^{-3/4}$ (the relation $\xi({\bar{\phi}}_{L})\simeq
D$ is not valid), the osmotic pressure $\Pi=\bar{\phi}_{L}^2
\partial_{\bar{\phi}_{L}} F_{c}/(N a^{3})$ (de Gennes, 1979) can be
derived in the form Eq.~2.

\vspace{5mm}
\noindent
{\bf Relation to bulk solution behavior:}
\noindent
For free polymer in solution, $\bar{\phi}_{L}\rightarrow \phi$, and
the free energy contains no
elastic restoring term.  From the blob expresion $F_{os}\simeq
k_{B}TN\phi^{5/4}$
we readily infer the des Cloizeaux expression Eq.~1 for bulk
polymers in the semi-dilute regime.  It is important to note that the
validity of Eq.~1 for bulk polymers is a necessary condition for the
validity of Eq.~2 for the compressed brush.  {\it In order to invoke
Eq.~2, it must be established that the polymer size and density in
the compressed brush are in a regime where bulk des Cloizeaux scaling
applies.}

%-------------------------------------------------------------------------
%Bibliography

\newpage
\vspace{5mm}
\noindent
{\bf REFERENCES}

\vspace{5mm}
\noindent
Alexander, S. 1977. Adsorption of chain molecules with a
polar head: a scaling description. {\it J.~Physique (Paris)}.
38:983-987.

\vspace{5mm}
\noindent
Auroy, P., L. Auvray, and L. Leger. 1991. Structures of
end-grafted polymer layers: a small-angle neutron scattering
study. {\it Macromolecules}. 24:2523-2528.

\vspace{5mm}
\noindent
Belsito, S., R. Bartucci, G. Montesano, D. Marsh, and L.
Sportelli. 2000.  Molecular and mesoscopic properties of hydrophilic
polymer-grafted
phospholipids mixed with phosphatidylcholine in aqueous dispersion: interaction
of dipalmitoyl N-poly(ethylene glycol)phosphatidylethanolamine with
dipalmitoylphosphatidylcholine studied by spectrophotometry and
spin-label electron spin resonance. {\it Biophys. J.} 78:1420-1430.

\vspace{5mm}
\noindent
Cohen, J., and S. Highsmith. 1997. An improved fit to
website osmotic pressure data. {\it Biophys. J.} 73:1689-1692.

\vspace{3mm}
\noindent
de Gennes, P.G. 1979. Scaling concepts in
polymer physics. Cornell University Press, Ithaca, New York.

\vspace{5mm}
\noindent
de Gennes, P.G. 1987. Polymers at an interface: a
simplified view. {\it Adv. Colloid Interface Sci.} 27:189-209.

\vspace{5mm}
\noindent
Dumitriu, S., editor. 2002. Polymeric Biomaterials. Marcel Dekker, New York.

\vspace{5mm}
\noindent
Evans, E., D.J. Klingenberg, W. Rawicz, and F. Szoka. 1996.
Interactions between polymer-grafted membranes in concentrated
solutions of free polymer. {\it Langmuir} 12:3031-3037.

\vspace{5mm}
\noindent
Hasse, H., H.-P. Kany, R. Tintinger, and G. Maurer.
1995. Osmotic virial coefficients of aqueous poly(ethylene
glycol) from laser-light scattering and isopiestic
measurements. {\it Macromolecules.} 28:3540-3552.

\vspace{5mm}
\noindent
Hristova, K., and D. Needham. 1995. Phase behavior of
a lipid/polymer-lipid mixture in aqueous medium. {\it Macromolecules.}
28:991-1002.

\vspace{5mm}
\noindent
Kenworthy, A.K., K. Hristova, D.~Needham, and
T.J. McIntosh. 1995a. Range and magnitude of the steric
pressure between bilayers containing phospholipids with
covalently attached poly(ethylene glycol). {\it Biophys. J.}
68:1921-1936.

\vspace{5mm}
\noindent
Kenworthy, A.K., S.A. Simon, and T.J.
McIntosh. 1995b. Structure and phase behavior of lipid
suspensions containing
phospholipids with covalently attached poly(ethylene
glycol). {\it Biophys. J.} 68:1903-1920.

\vspace{5mm}
\noindent
Kuhl, T.L., D.E.~Leckband,  D.D. Lasic, and J.N.
Israelachvili. 1994. Modulation of interaction forces between
bilayers exposing short-chained ethylene oxide headgroups.
{\it Biophys. J.} 66:1479-1488.

\vspace{5mm}
\noindent
Lasic, D.D. 1997. The conformation of polymers at
interfaces. {\it In} Poly(ethylene glycol) Chemistry and
Biological Applications. J.M. Harris and S. Zalipsky,
editors. American Chemical Society, Washington, DC. 31-44.

\vspace{5mm}
\noindent
Lasic, D.D., and F. Martin, editors. 1995. Stealth Liposomes.
CRC Press, Boca Raton.

\vspace{5mm}
\noindent
Lasic, D.D., and D. Papahadjopoulos, editors. 1998. Medical Applications
of Liposomes. Elsevier Science, Amsterdam.

\vspace{5mm}
\noindent
Lis, L.J., M. McAlister, N. Fuller, P. Rand, and V.A. Parsegian. 1982.
Interactions between neutral phospholipid bilayer membranes.
{\it Biophys J.} 37:657-666.

\vspace{5mm}
\noindent
Milner, S.T., T.A. Witten, and M.E. Cates. 1988. Theory of the
grafted polymer brush. {\it Macromolecules.} 21:2610-2619.

\vspace{5mm}
\noindent
Montesano, G., R. Bartucci, S. Belsito, D. Marsh, and L. Sportelli. 2001.
Lipid membrane expansion and micelle formation by
polymer-grafted lipids: scaling
with polymer length studied by spin-label electron spin resonance.
{\it Biophys. J.} 80:1372-1383.

\vspace{5mm}
\noindent
Needham, D., K. Hristova, T.J. McIntosh, M. Dewhirst,
N. Wu, and D.D. Lasic. 1992. Polymer-grafted
liposomes: physical basis for the ``stealth'' property.
{\it J. Liposome Res.} 2:411-430.

\vspace{5mm}
\noindent
Rand, P.R. 2002. http://aqueous.labs.brocku.ca/osfile.html.

\vspace{5mm}
\noindent
Rand, P.R., and V.A. Parsegian. 1989. Hydration forces
between phospholipid bilayers. {\it Biochim. Biophys.
Acta.} 988:351-376.

\vspace{5mm}
\noindent
Reid, C., and R.P. Rand. 1997. Fits to osmotic pressure data. {\it 
Biophys. J.} 73:1692-1694.

\vspace{5mm}
\noindent
Sackett, D.L. 1995. Structure and function in the tubulin dimer and
the role of the acidic carboxyl terminus. {\it In} Subcellular
Biochemistry, Volume 24. Proteins: Structure, Function, and
Engineering. B.B. Biswas and S. Roy, editors. Plenum Press,
New York. 255-302.

\vspace{5mm}
\noindent
Szleifer, I. 1996. Statistical thermodynamics of polymers near
surfaces. {\it Curr. Opin. Coll. Int. Sci.} 1:416-423.

\vspace{5mm}
\noindent
Szleifer, I. 1997. Polymers and proteins: interactions at interfaces.
{\it Curr. Opin. Solid State Mat. Sci.} 2:337-344.

\vspace{5mm}
\noindent
Taunton, H.J., C. Toprakcioglu, L.J. Fetters, and
J. Klein. 1990. Interactions between surfaces bearing end-adsorbed
chains in a good solvent. {\it Macromolecules.} 23:571-580.
%-------------------------------------------------------------------------
% Figure Captions

\newpage
\noindent

\vspace{1cm}
\noindent
{\bf Figure 1:} Schematic drawing of two apposed brush layers of 
surface-grafted PEG polymers in a multilamellar liposome.  $L_0$ is the 
unperturbed thickness of the polymer brush, $d_f$ is the surface-to-surface 
distance, $L \equiv {d_f/2}$, and $D$ is the distance 
between grafting sites.  Osmotic stress forces the surfaces together, leading 
to interaction and interpenetration of 
the brushes such that $L < L_0$.  $\xi$ is the blob size of the polymer.
The yellow regions denote the lipid bilayers, and the blue region denotes 
the aqueous solution in between lamellae of the liposome. 

\vspace{1cm}
\noindent
{\bf Figure 2:} Room-temperature bulk osmotic pressure $\Pi$ of various PEG
polymers vs. monomer volume fraction $\phi$.  The symbols are data 
obtained from Rand (2002) for PEGs of molecular weights 
$1000,1500,2000,4000,
8000,10000$ and $20000$ Da.  The monomer volume fraction is
$\phi\simeq 0.59\times w$, where $0 <w <0.5$ is the weight
fraction of polymer (Rand, 2002).  (In obtaining the relation
between $\phi$ and $w$ we have assumed the specific density of
PEG solutions to be $1$ g/cm$^{3}$ (Hasse et al., 1995).)
The solid line is the des Cloizeaux equation $\Pi=\alpha (k_{B}T/a^{3})
\phi^{9/4}$, with $a=3.5$ {\AA} (Kenworthy et al., 1995a) and $\alpha = 0.8$.
Note that at high $\phi$ the data converge to a universal straight
line independent of molecular weight, in agreement with the des Cloizeaux
prediction.  For reference, the vertical dashed lines are the
molecular-weight-independent monomer volume fractions in an 
uncompressed brush $ \bar{\phi}_{0}={(a/D)}^{4/3}$ for
PEG-lipid mol fractions $f= 0.05, 0.1,$ and $0.2$ if
$a=3.5$ {\AA} and $D={(A/f)}^{1/2}$, where $A=48$ \AA$^{2}$
(Kenworthy et al., 1995a).  A surface-grafted PEG layer is in the brush
scaling regime if its free-polymer bulk $\Pi$ lies on the des Cloizeaux 
line at the monomer volume-fraction $\phi$ present in the surface 
layer (vertical dashed lines).  Thus, at a 10\% grafting density, a PEG-6000 
layer is in the brush regime, but a PEG-1500 layer is not.  (See text and Appendix for details.)

%\newpage
\vspace{3mm}
\noindent
{\bf Figure 3:} Osmotic pressure $\Pi$ vs. fluid spacing $d_{f}=2L$
between apposing PEG-grafted bilayers.
The symbols are data from Kenworthy et al. (1995a) for Avanti
DSPC:PEG-5000 complexes with nominal
grafting mol fractions $f=0, 0.015, 0.03, 0.05, 0.1,$ and $0.2$.
The solid curves are fits to the $f=0.1$ and $0.2$ data
using Eqs.~2 and 3.
Two-parameter unconstrained fits
of the AdG prediction for osmotic pressure vs. fluid spacing yield
nearly constant effective monomer lengths $a_{0.1}=3.56\pm 0.07$ {\AA}
and $a_{0.2}=3.30\pm 0.15$
{\AA}.  These values agree well with structural predictions reported in the
literature (Kenworthy et al., 1995a), lending veracity to the
applicability of AdG theory
to these data.  The $f=0.015, 0.03, 0.05$ data do not meet the brush
criterion (see Fig.~2) and thus AdG theory is not applicable.
(See text for details.)

\newpage

\begin{figure}[h]
\begin{center}
\epsfig{file=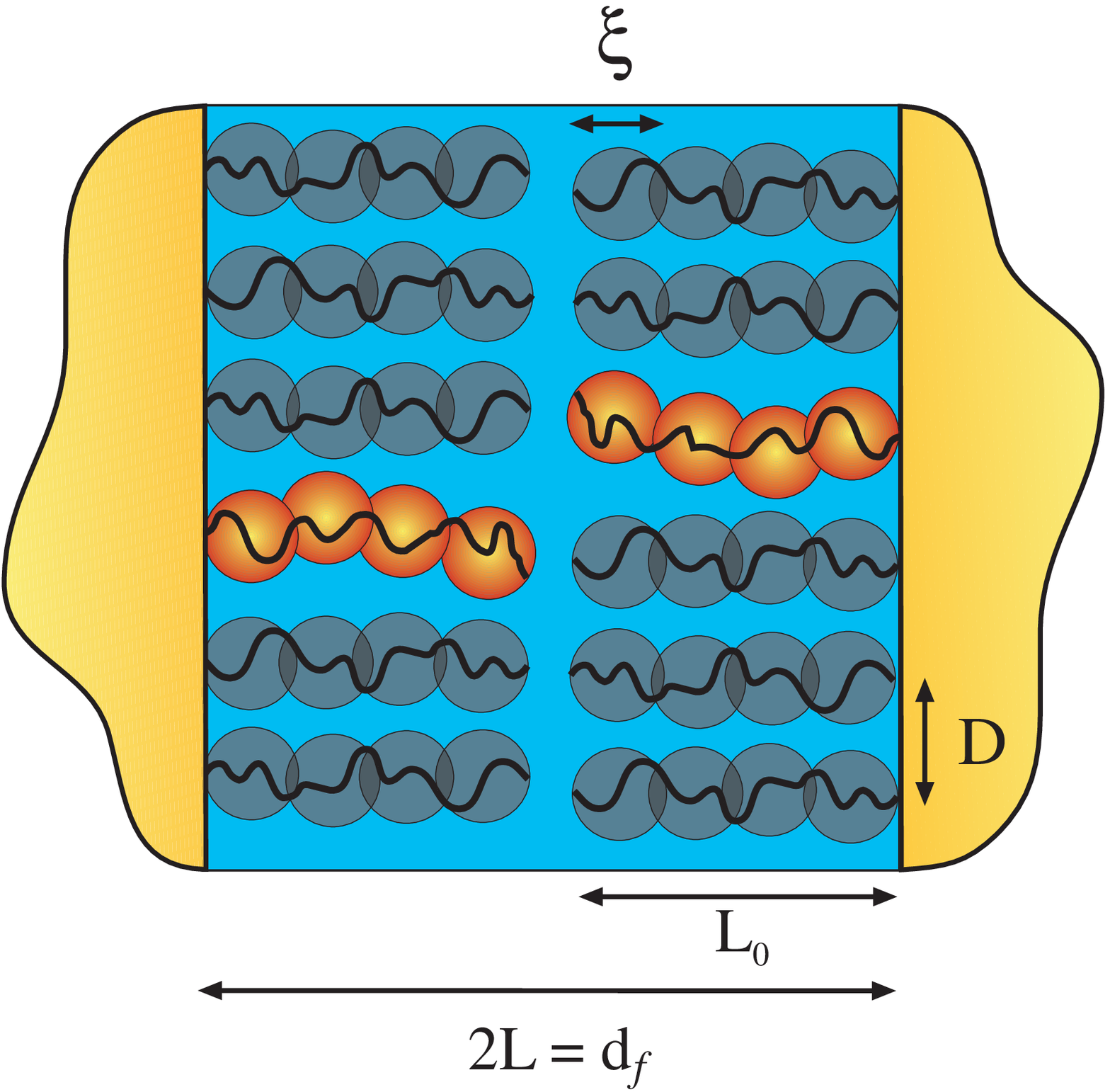, width=3.0in, angle=0}
\end{center}
\caption{~}
\label{figure1}
\end{figure}

\begin{figure}[h]
\begin{center}
\epsfig{file=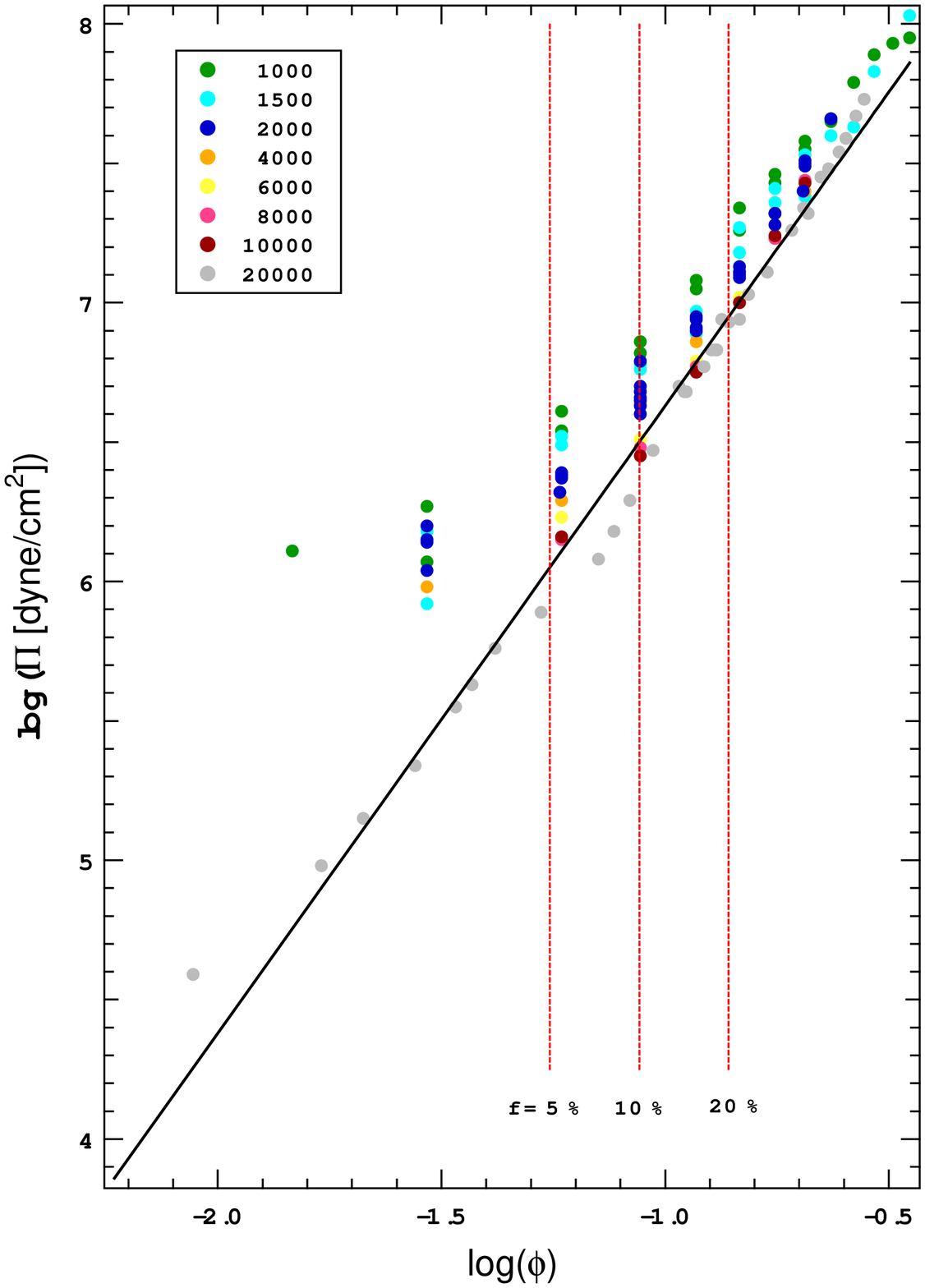, width=4.0in, angle=0}
\end{center}
\caption{~}
\label{figure2}
\end{figure}

\begin{figure}[h]
\begin{center}
\epsfig{file=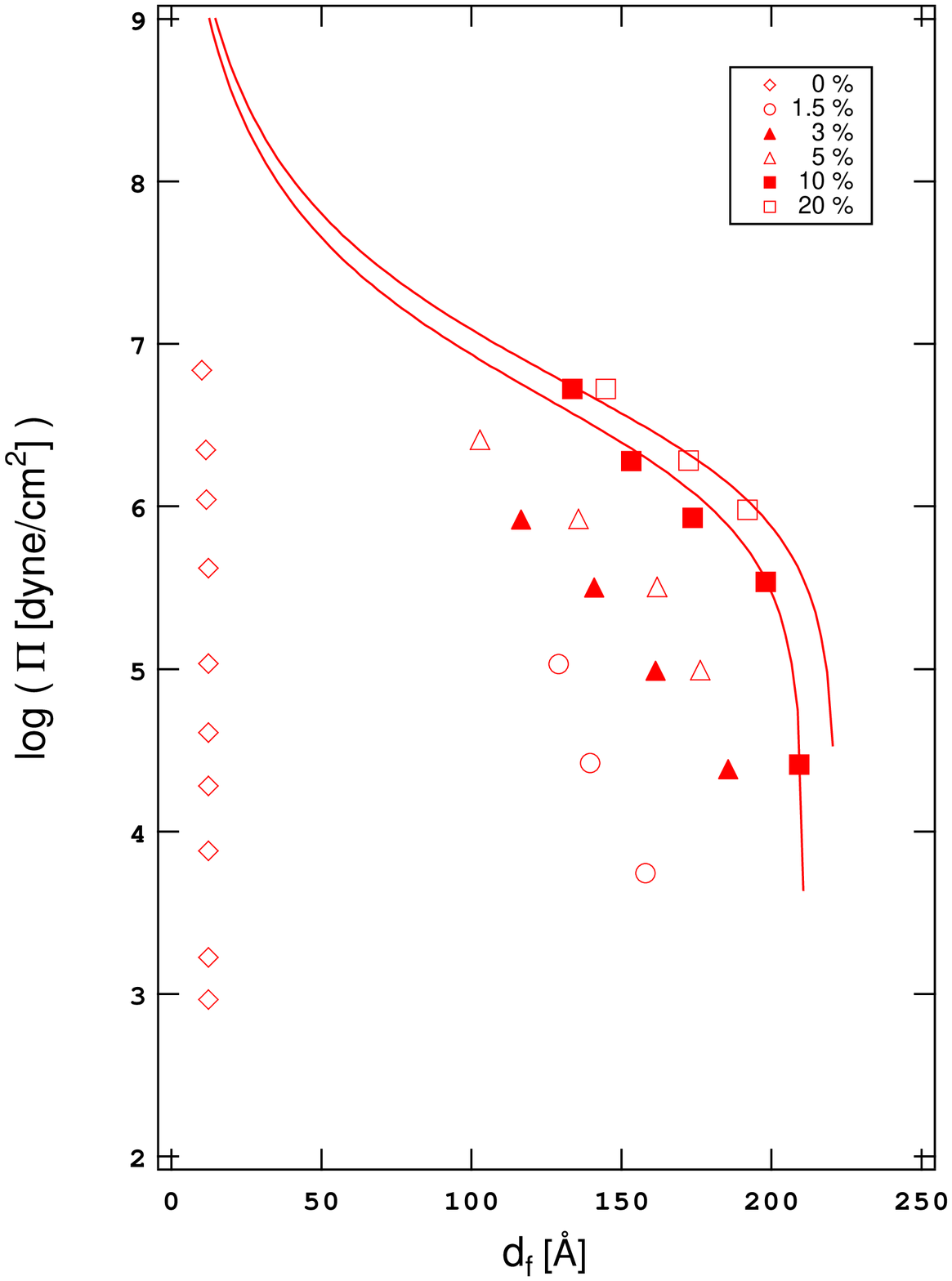, width=4.5in, angle=0}
\end{center}
\caption{~}
\label{figure3}
\end{figure}

\end{document}